\documentclass[letterpaper, 10 pt, conference]{ieeeconf}  

\IEEEoverridecommandlockouts                              

\usepackage{graphicx}

\usepackage{algorithm}
\usepackage{algpseudocode}

\usepackage{amsmath} 
\usepackage{amssymb}  
\usepackage{cite}

\usepackage{subcaption}  

\newtheorem{definition}{Definition}
\newtheorem{proposition}{Proposition}
\newtheorem{assumption}{Assumption}
\newtheorem{remark}{Remark}

\title{\LARGE \bf
Distribution-Free Probabilistic Environmental Control Barrier Functions with Application to Adaptive Cruise Control
}

\author{Byeongyu Lee$^{1}$, Kyoungseok Han$^{1}$,  Chung Choo Chung$^{2}$
\thanks{$^{1}$Byeongyu Lee and Kyoungseok Han are with the Department of Automotive Engineering (Automotive-Computer Convergence), Hanyang University, Seoul 04763, Republic of Korea, Correspondence to: {\tt\small dlquddb@hanyang.ac.kr, kyoungsh@hanyang.ac.kr}}
\thanks{$^{2}$Chung Choo Chung is with the Department of Electrical Engineering,
Hanyang University, Seoul 04763, Republic of Korea, Correspondence to: {\tt\small cchung@hanyang.ac.kr}}
}

\begin{document}

\maketitle
\thispagestyle{empty}
\pagestyle{empty}

\begin{abstract}
Safety control under environmental uncertainty is challenging
when the uncertainty distribution is unknown and reliable support bounds are unavailable
This paper introduces CPP-ECBF, a distribution-free approach
that combines conformal predictive programming with
probabilistic environmental control barrier functions.
For separable environmental safety functions and i.i.d.
uncertainty, CPP-ECBF converts the probabilistic safety
condition into a sample-based deterministic control constraint
that provides finite-horizon safety with prescribed probability
$1-\epsilon$ and calibration confidence
$1-\widetilde{\beta}$.
For adaptive cruise control, an offline conformal uncertainty
bound yields an explicit linear upper bound on the ego-vehicle
acceleration, which is enforced by a quadratic-programming
safety filter.
In $10{,}000$ Monte Carlo trials, $99.97\%$ of trials satisfied the
finite-horizon safety criterion, exceeding the prescribed
confidence of $85\%$.
Moreover, CPP-ECBF reduced the conservatism of worst-case
robust enforcement, increasing the admissible acceleration
upper limit by $5.90\,\mathrm{m/s^2}$ on average.
\end{abstract}

\section{INTRODUCTION}

Control Barrier Functions (CBFs) provide a systematic means of
enforcing safety constraints~\cite{CBF}, with applications such as
Adaptive Cruise Control (ACC)~\cite{CBF_ACC}.
Environmental Control Barrier Functions (ECBFs) extend this
approach to safety functions involving environmental states and
provide robust safety conditions under bounded environmental
uncertainty~\cite{ECBF}.
However, reliable uncertainty bounds may be unavailable, and
worst-case enforcement can be conservative even when valid bounds
are known. This motivates a probabilistic treatment of environmental
uncertainty, whose underlying probability distribution is often unknown.
Probabilistic CBFs provide prescribed finite-horizon safety
guarantees for discrete-time stochastic systems by imposing
chance constraints on the control input~\cite{PCBF}.
The scenario approach (SA) has been used to enforce these
constraints using uncertainty samples when the underlying
distribution is unknown~\cite{PCBF}.
In addition to SA, sample average approximation (SAA) is
widely used for chance-constrained optimization (CCO) under
unknown distributions~\cite{SA,SAA}.
Standard SA guarantees rely on convexity, while finite-sample
feasibility guarantees for SAA require additional
conditions~\cite{CPP}.

Conformal prediction (CP) constructs prediction sets with
finite-sample, distribution-free coverage guarantees using a
calibration dataset, without requiring a parametric model of the
underlying uncertainty distribution~\cite{CP_tutorial,CP}.
Conformal predictive programming (CPP) leverages these coverage
guarantees to address chance-constrained optimization (CCO)
problems~\cite{CPP}.
For general chance constraints, coupling between the decision
variable and uncertainty can prevent CPP from guaranteeing
the prescribed probability~\cite{CPP}.
For separable constraints, however, the uncertain component is
independent of the decision variable, allowing CPP to provide
a finite-sample guarantee at the prescribed probability~\cite{CPP}.
CP has also been used for safety-critical control in dynamic
environments~\cite{CP_MPC} and under perception
uncertainty~\cite{CP_Perception}.
However, translating their conformal coverage guarantees into
safety guarantees requires additional problem-specific
assumptions, such as Lipschitz
continuity~\cite{CP_MPC,CP_Perception}.

In this work, we introduce CPP-ECBF and derive an explicit
deterministic linear safety constraint for ACC
under environmental uncertainty with an unknown distribution.
The constraint is obtained by applying CPP to a probabilistic
ECBF condition under a separable environmental safety structure.
We show that enforcing the resulting constraint 
guarantees the prescribed finite-horizon safety probability with
prescribed confidence over the calibration dataset under the stated
i.i.d.\ sampling assumptions.
By permitting a prescribed violation probability, the
constraint can alleviate worst-case conservatism.
For the considered ACC problem, the lead-vehicle acceleration
is modeled as exogenous environmental uncertainty and the safety
function separates into ego-vehicle and lead-vehicle components.
This structure allows an offline conformal uncertainty quantile
to be computed directly from lead-vehicle acceleration samples.
Substituting this bound into the CPP-ECBF condition yields an
explicit linear upper bound on the ego-vehicle acceleration,
which a quadratic-programming (QP) safety
filter enforces.
Monte Carlo simulations empirically assess the finite-horizon safety
performance, compare conditional and marginal CPP-ECBF, and
quantify the reduction in conservatism relative to
worst-case robust enforcement.

\section{Problem Setting and Preliminaries}

\subsection{Notation}
For any indexed sequence, let us denote $ a_{m:n}:= a_m,\ldots,a_n$ and $a^{(m:n)}:=a^{(m)},\ldots,a^{(n)}, \ \ m,n\in \mathbb{N}, \ m\leq n.$ 
Given $\delta \in (0,1)$ and a collection of $N \in \mathbb{N}$ real scalars $R^{(1:N)}$,
$\mathrm{Quantile}_{1-\delta}(R^{(1:N)}, \infty)$ denotes the $(1-\delta)$-quantile of the collection,
that is, the value obtained by sorting the elements in non-decreasing order
and selecting the $p$-th one, where $p = \lceil (N+1)(1-\delta) \rceil$ and $\lceil \cdot \rceil$ denotes the ceiling operator. Let us denote $\mathcal{C}_{1-\delta}(R^{(1:N)}) := \mathrm{Quantile}_{1-\delta}\big(R^{(1:N)},\infty\big).$
%

\subsection{Problem Setting}

Let us consider a discrete-time system given by
\begin{equation}
    \mathbf{x}_{k+1}=F(\mathbf{x}_k,\mathbf{u}_k),
    \label{eq:system}
\end{equation}
where $\mathbf{x}_k\in\mathcal{X}\subset\mathbb{R}^n$
is the state, $\mathbf{u}_k\in\mathcal{U}\subset\mathbb{R}^m$
is the control input, and $\mathcal{X}$ and $\mathcal{U}$
are the sets of feasible states and control inputs,
respectively.
Safety also depends on an environmental state $\mathbf{e}_k\in\mathbb{R}^{n_e}$, which evolves as
\begin{equation}
    \mathbf{e}_{k+1}=F_e(\mathbf{e}_k,\mathbf{d}_k),
    \label{eq:e_system}
\end{equation}
where $\mathbf{d}_k\in\mathbb{R}^{n_d}$ represents environmental
uncertainty.
The distribution of $\mathbf d_k$ is assumed to be unknown,
and no known lower or upper bounds on $\mathbf d_k$ are required.
Let us define the environment-dependent safe set $\mathcal{S}_e$
as the $0$-superlevel set of a continuous safety function
$h_e(\mathbf{x},\mathbf{e})$.
\[
\mathcal{S}_e
:=
\left\{
(\mathbf{x},\mathbf{e})\in\mathcal{X}\times\mathbb{R}^{n_e}
\mid
h_e(\mathbf{x},\mathbf{e})\geq0
\right\},
\]
where
$h_e:\mathcal{X}\times\mathbb{R}^{n_e}\rightarrow\mathbb{R}$
is continuous.
For a
user-specified horizon $H\in\mathbb{N}$ and failure probability
$\epsilon\in(0,1)$, the objective is to ensure that, starting from
any $(\mathbf{x}_k,\mathbf{e}_k)\in\mathcal{S}_e$, safety is
maintained over the horizon with probability at least
$1-\epsilon$, i.e.,
\begin{equation}
\mathbb{P}_{\mathbf{d}_{k:k+H-1}}\!\left(
\bigcap_{i=1}^{H}
\{(\mathbf{x}_{k+i},\mathbf{e}_{k+i})\in\mathcal{S}_e\}
\right)
\geq 1-\epsilon.
\label{eq:finite_horizon_safety_goal}
\end{equation}
Equation \eqref{eq:finite_horizon_safety_goal} is referred to as
\emph{$\epsilon$-safety over horizon $H$} that allows a prescribed probability of safety
violation over a finite horizon, rather than requiring the environment-dependent safe set
$\mathcal{S}_e$ to remain forward invariant for every realization of the
environmental uncertainty. 
Our goal is to derive a sample-based control input constraint that can
be enforced at each time step to guarantee
\eqref{eq:finite_horizon_safety_goal}.

\subsection{Conformal Prediction}

CP constructs prediction sets with finite-sample
coverage guarantees without requiring the underlying probability
distribution to be known~\cite{CP_tutorial,CP}. Given an uncertain
outcome $Y$ and its prediction $\widehat{Y}$, CP uses a scalar score
$R:=\ell(\widehat{Y},Y)$, where $\ell$ is a user-defined score function.
The quantities $R^{(1:N)}$ and $R^{\mathrm{test}}$ are referred to as \emph{calibration scores} and \emph{test score}, respectively.
CP provides a finite-sample probabilistic guarantee relating the
calibration scores to the test score under the following assumption.

\begin{assumption}
\label{a:iid}
The calibration scores
$R^{(1)},\ldots,R^{(N)}$ and the test score
$R^{\mathrm{test}}$ are independent and identically distributed.
\end{assumption}

Under Assumption~\ref{a:iid}, CP provides a
marginal coverage guarantee of the form~\cite{CP_tutorial}
\[
\mathbb{P}_{R^{(1:N)},R^{\mathrm{test}}}\!\left(
R^{\mathrm{test}}
\leq
\mathcal{C}_{1-\delta}(R^{(1:N)})
\right)
\geq 1-\delta.
\]
This marginal guarantee is over both the calibration scores
and the test score. In this paper, a single calibration dataset is held fixed during closed-loop operation; hence, a calibration-conditional guarantee is used. 
For simplicity of notation, let us denote
\[
\widetilde{\mathcal C}(R^{(1:N)})
:=
\mathcal C_{
1-\delta+\sqrt{\frac{\ln(1/\beta)}{2N}}
}
(R^{(1:N)}).
\]
The following proposition gives a conditional CP guarantee.

\begin{proposition}[Conditional CP]
\label{p:conditional_CP}
Let $\delta, \beta \in (0,1)$ and $\sqrt{\frac{\ln(1/\beta)}{2N}}<\delta$. Under Assumption~\ref{a:iid}, it holds that
\[
\mathbb{P}_{ R^{(1:N)}}\Big(
\mathbb{P}_{R^{\text{test}
}}\!\big(
R^{\text{test}} \le  \widetilde{\mathcal{C}}(R^{(1:N)}) 
\,\big|\, R^{(1:N)}
\big) \ge 1-\delta
\Big) \ge 1-\beta 
\]
\end{proposition}

Here, the inner probability is the conditional coverage for
a fixed calibration dataset. The outer inequality guarantees
that this conditional coverage is at least $1-\delta$ with
probability at least $1-\beta$ over calibration datasets. $1-\delta$ and $1-\beta$ are referred as the
prescribed coverage and prescribed confidence, respectively.

$Proof.$ The detailed proof follows the argument in~\cite{CP}
and is omitted for brevity.
By Hoeffding's inequality, increasing the quantile
level by $\sqrt{\ln(1/\beta)/(2N)}$ ensures
conditional coverage of at least $1-\delta$
with confidence at least $1-\beta$ over the
calibration dataset.
\hfill$\square$

\subsection{Conformal Predictive Programming}

Conformal predictive programming (CPP) applies CP to CCO problems~\cite{CPP}. Let us consider the chance constraint
\[
\mathbb{P}_{\mathbf{b}^{test}}\!\big(f(\mathbf{a},\mathbf{b}^{test})\leq 0\big) \ge 1-\delta,
\]
where $\mathbf{a}\in \mathbb{R}^q$ is a decision variable and $\mathbf{b}\in \mathbb{R}^r$ is a random variable. Let $\mathbf{b}^{(1:N)}$ denote a calibration dataset and assume
that
$\mathbf{b}^{(1:N)},\mathbf{b}^{\mathrm{test}}
\overset{\mathrm{i.i.d.}}{\sim}P_{\mathbf b}$.
The following proposition establishes a probabilistic guarantee.

\begin{proposition}[CPP]
\label{p:CPP_Guarantee}
Let $\delta, \beta \in (0,1)$. Let  $f(\mathbf{a}, \mathbf{b})$ be of the form $f(\mathbf{a}, \mathbf{b}):=\phi(\mathbf{a}) - \psi(\mathbf{b})$ where $\phi:\mathbb{R}^q\to \mathbb{R}$ and $\psi : \mathbb{R}^r\to \mathbb{R}$ are arbitrary functions. 
Then, it holds that
\[
\mathbb{P}_{\mathbf{b}^{\text{test}},\,\mathbf{b}^{(1:N)}}
\Big(
f(\mathbf{a}, \mathbf{b}^{\text{test}})
\le
\mathcal{C}_{1-\delta}\!\left(f^{(1:N)}\right)
\Big)
\ge 1-\delta.
\]
Furthermore, if $\sqrt{\frac{\ln(1/\beta)}{2N}}< \delta$, then
\begin{equation}
\begin{aligned}
\mathbb{P}_{\mathbf{b}^{(1:N)}}\!\Bigg(
\mathbb{P}_{\mathbf{b}^{\mathrm{test}}}\!\Big(
f(\mathbf{a},\mathbf{b}^{\mathrm{test}})
&\leq
\widetilde{\mathcal{C}}\!\left(f^{(1:N)}\right)
\,\Big|\,\mathbf{b}^{(1:N)}
\Big) \\
&\geq 1-\delta
\Bigg)
\geq 1-\beta,
\end{aligned}
\label{eq:CPP_ineq}
\end{equation}
where
\[
f^{(i)}:=f(\mathbf{a}, \mathbf{b}^{(i)}).
\]
\end{proposition}
$Proof.$ The proof follows the argument in~\cite{CPP}
and is omitted for brevity.

\section{Probabilistic Environmental Control Barrier Functions}

Hereafter, all CBF conditions are written in discrete-time
exponential form, and the qualifiers ``discrete-time'' and
``exponential'' are omitted for brevity.
This section first introduces a one-step probabilistic ECBF
condition and relates it to the prescribed finite-horizon
safety requirement.
For separable environmental safety functions, CPP is then used
to derive a sample-based control input constraint.
We prove that this constraint satisfies the probabilistic ECBF
condition and thereby guarantees the prescribed finite-horizon
safety property.

\subsection{Probabilistic ECBF}

For notational convenience, let us denote
\[
\begin{aligned}
&\Delta h_e(\mathbf{x}_k,\mathbf{e}_k,\mathbf{u}_k,\mathbf{d}_k)\\
& :=
h_e\!\left(
F(\mathbf{x}_k,\mathbf{u}_k),
F_e(\mathbf{e}_k,\mathbf{d}_k)
\right)
-\alpha h_e(\mathbf{x}_k,\mathbf{e}_k),
\ \ \alpha\in[0,1).
\end{aligned}
\]
Following the probabilistic CBF formulation
in~\cite{PCBF}, the probabilistic ECBF condition is defined through the following calibration-conditional chance constraint.
Given a probability distribution $P_{\mathbf d}$, let us define {\it probabilistic ECBF}.
\begin{definition}[probabilistic ECBF]
Let $\delta,\beta\in(0,1)$ and
$\mathbf{d}^{(1:N)},\mathbf{d}_k\overset{\mathrm{i.i.d.}}{\sim}P_{\mathbf d}$.
A function
$h_e:\mathcal{X}\times\mathbb{R}^{n_e}\to\mathbb{R}$
is a $(\delta,\beta)$-probabilistic ECBF for systems
\eqref{eq:system} and \eqref{eq:e_system} if
\begin{enumerate}
\item for every $(\mathbf{x}_k,\mathbf{e}_k)\in\mathcal{S}_e$ and
\item there exists a control input $\mathbf{u}_k\in\mathcal{U}$ such that
\[
\begin{aligned}
\mathbb{P}_{\mathbf{d}^{(1:N)}}\!\Bigg(
\mathbb{P}_{\mathbf{d}_k}\!\Big(
\Delta h_e( \mathbf{x}_k,\mathbf{e}_k,\mathbf{u}_k,\mathbf{d}_k)
&\geq0
\,\Big|\,\mathbf{d}^{(1:N)}
\Big) \\
&\geq1-\delta
\Bigg)
\geq1-\beta.
\end{aligned}
\]
\end{enumerate}
\label{d:PECBF}
\end{definition}
\begin{remark}
In the proposed safety-filtering scheme, the control input is
selected subject to a constraint that depends on the fixed
calibration dataset and the prescribed probability parameters,
as shown explicitly in Section~III-B. Hence, more explicitly,
the resulting control input may be written as
\[
\mathbf{u}_k
=
\pi\!\left(
\mathbf{x}_k,\mathbf{e}_k;
\mathbf{d}^{(1:N)},\delta,\beta
\right).
\]
Accordingly, the inner probability in
Definition~\ref{d:PECBF} is conditioned on the fixed calibration
dataset $\mathbf{d}^{(1:N)}$. For notational simplicity, we suppress this explicit dependence hereafter and denote the control input simply by $\mathbf{u}_k$.
\end{remark}

The following proposition relates this one-step safety condition to the
finite-horizon $\epsilon$-safety requirement in
\eqref{eq:finite_horizon_safety_goal}.

\begin{proposition}[probabilistic ECBF]
\label{p:PECBF}
Let $\epsilon,\widetilde{\beta}\in(0,1)$ and
$\mathbf{d}^{(1:N)}, \mathbf{d}_k\overset{\mathrm{i.i.d.}}{\sim}P_{\mathbf d}$.
Further, assume that $\delta,\beta\in(0,1)$ satisfy
\begin{equation}
\delta \leq 1-(1-\epsilon)^{\frac{1}{H}}
\qquad \text{and} \qquad
\beta \leq \frac{\widetilde{\beta}}{H}.
\label{eq:PECBF_ineq}
\end{equation}
Let $h_e$ be a $(\delta,\beta)$-probabilistic ECBF for systems
\eqref{eq:system} and \eqref{eq:e_system}.
If $(\mathbf{x}_k,\mathbf{e}_k)\in\mathcal{S}_e$ and the control input $\mathbf{u}_k$ is chosen at each time step such that
\begin{equation}
\begin{aligned}
\mathbb{P}_{\mathbf{d}^{(1:N)}}\!\Bigg(
\mathbb{P}_{\mathbf{d}_k}\!\Big(
\Delta h_e( \mathbf{x}_k,\mathbf{e}_k,\mathbf{u}_k,\mathbf{d}_k)
&\geq0
\,\Big|\,\mathbf{d}^{(1:N)}
\Big) \\
&\geq1-\delta
\Bigg)
\geq1-\beta,
\end{aligned}
\label{eq:PECBF_uineq}
\end{equation}
then the environment-dependent safe set $\mathcal{S}_e$ is
$\epsilon$-safe over the time horizon $H$ with confidence
$1-\widetilde{\beta}$, i.e.,
\[
\begin{aligned}
\mathbb{P}_{\mathbf{d}^{(1:N)}}\!\Bigg(
\mathbb{P}_{\mathbf{d}_{k:k+H-1}}\!\Bigg(
\bigcap_{i=1}^{H}
\{(\mathbf{x}_{k+i},& \mathbf{e}_{k+i})\in\mathcal{S}_e\} 
\,\Bigg|\,\mathbf{d}^{(1:N)}
\Bigg) \\
&\geq1-\epsilon
\Bigg)
\geq1-\widetilde{\beta}.
\end{aligned}
\label{eq:PECBF_goal}
\]
\end{proposition}
$Proof.$ The detailed proof is omitted for brevity.
Following the argument in~\cite{PCBF}, the one-step probabilistic
ECBF condition yields an $H$-step safety probability of at least
$(1-\delta)^H$.
Hence, $(1-\delta)^H\geq1-\epsilon$ is sufficient for the
prescribed finite-horizon safety probability.
Moreover, applying Fr\'echet's inequality to the $H$ confidence
events gives an overall confidence of at least $1-H\beta$.
Thus, $1-H\beta\geq1-\widetilde{\beta}$ is sufficient for the
prescribed confidence. Rearranging these two conditions gives
\eqref{eq:PECBF_ineq}.
\hfill$\square$

Thus, for a prescribed horizon $H$, failure probability $\epsilon$,
and confidence $1-\widetilde{\beta}$, choosing $\delta$ and $\beta$
according to \eqref{eq:PECBF_ineq} allows the one-step probabilistic
ECBF condition to guarantee $\epsilon$-safety over horizon $H$ with
confidence $1-\widetilde{\beta}$.

\subsection{CPP-ECBF}

For a separable environmental safety function, the uncertain
component does not depend on the control input, allowing CPP to
construct a sample-based control input constraint.

\begin{definition}[CPP-ECBF]
Let $\delta,\beta\in(0,1)$ and
$\mathbf{d}^{(1:N)},\mathbf{d}_k
\overset{\mathrm{i.i.d.}}{\sim}P_{\mathbf d}$.
A function
$h_e:\mathcal{X}\times\mathbb{R}^{n_e}\to\mathbb{R}$
is a $(\delta,\beta)$-CPP-ECBF for systems
\eqref{eq:system} and \eqref{eq:e_system} if
\begin{enumerate}
\item the safety function has the separable form
\[
h_e(\mathbf{x},\mathbf{e})
=
\phi(\mathbf{x})-\psi(\mathbf{e}),
\]
where $\phi:\mathcal{X}\to\mathbb{R}$ and
$\psi:\mathbb{R}^{n_e}\to\mathbb{R}$,
\item for every $(\mathbf{x}_k,\mathbf{e}_k)\in\mathcal{S}_e$ and
\item there exists a control input
$\mathbf{u}_k\in\mathcal{U}$ such that
\[
\phi\!\big(F(\mathbf{x}_k,\mathbf{u}_k)\big)
\geq
\widetilde{\mathcal{C}}\!\left(\psi^{(1:N)}_k\right)
+
\alpha h_e(\mathbf{x}_k,\mathbf{e}_k), \ \ \alpha \in [0,1),   
\]
where
\[
\psi^{(i)}_k
:=
\psi\!\left(F_e(\mathbf{e}_k,\mathbf{d}^{(i)})\right).
\]
\end{enumerate}
\label{d:CPP-ECBF}
\end{definition}
%


Because the construction in Definition~\ref{d:CPP-ECBF}
uses the calibration-conditional conformal bound
$\widetilde{\mathcal C}$, we refer to it as
\emph{conditional CPP-ECBF} when distinguishing it from
the marginal variant considered in Section~IV.
For brevity, we simply use CPP-ECBF elsewhere unless
such a distinction is necessary.
The following proposition states the safety guarantee induced by
Definition~\ref{d:CPP-ECBF}.

\begin{proposition}[CPP-ECBF]
\label{p:last}
Let $\epsilon,\widetilde{\beta}\in(0,1)$ and
$\mathbf{d}^{(1:N)}
\overset{\mathrm{i.i.d.}}{\sim}P_{\mathbf d}$. Let
$\delta,\beta\in(0,1)$ satisfy \eqref{eq:PECBF_ineq} and
$
\sqrt{\frac{\ln(1/\beta)}{2N}}<\delta.
$
Let $h_e$ be a $(\delta,\beta)$-CPP-ECBF for systems
\eqref{eq:system} and \eqref{eq:e_system}. If $(\mathbf{x}_k,\mathbf{e}_k)\in\mathcal{S}_e$ and the control input $\mathbf{u}_k$  is chosen at each time step such that
\begin{equation}
\phi\!\big(F(\mathbf{x}_k,\mathbf{u}_k)\big)
\geq
\widetilde{\mathcal{C}}\!\left(\psi^{(1:N)}_k\right)
+
\alpha h_e(\mathbf{x}_k,\mathbf{e}_k),
\label{eq:u_policy}
\end{equation}
where
\[
\psi^{(i)}_k:= \psi(F_e(\mathbf{e}_k, \mathbf{d}^{(i)})),
\]
then the environment-dependent safe set $\mathcal{S}_e$ is
$\epsilon$-safe over the time horizon $H$ with confidence
$1-\widetilde{\beta}$.
\end{proposition}

$Proof.$ It is enough to show that
\eqref{eq:u_policy} implies \eqref{eq:PECBF_uineq}.
Expanding $\Delta h_e$ gives
\[
\begin{aligned}
&\Delta  h_e(\mathbf{x}_k,\mathbf{e}_k,\mathbf{u}_k,\mathbf{d}_k)  \\ 
& = \phi (F(\mathbf{x}_k, \mathbf{u}_k))-\psi(F_e(\mathbf{e}_k, \mathbf{d}_k))-\alpha (\phi (\mathbf{x}_k)-\psi(\mathbf{e}_k))
\end{aligned}
\]
Because $h_e$ is separable, $-\Delta h_e$ has the separable
form required by Proposition~\ref{p:CPP_Guarantee}.
For the calibration samples, 
let us denote
\[
\Delta h_e^{(i)}
:=
\Delta  h_e(\mathbf{x}_k,\mathbf{e}_k,\mathbf{u}_k,\mathbf{d}^{(i)}).
\]
Applying the conformal quantile to
$-\Delta h_e^{(1:N)}$ gives
\begin{equation}
\widetilde{\mathcal C}(-\Delta h_e^{(1:N)})
=
\widetilde{\mathcal C}(
-\phi(F(\mathbf{x}_k, \mathbf{u}_k))
+\psi_k^{(1:N)}
+\alpha h_e(\mathbf{x}_k,\mathbf{e}_k)).
\label{eq:11}
\end{equation}
Because the terms other than $\psi_k^{(1:N)}$ do not depend on the
calibration sample index, \eqref{eq:11} becomes
\[
\widetilde{\mathcal C}(-\Delta h_e^{(1:N)})
=
\widetilde{\mathcal C}(\psi_k^{(1:N)})
-\phi(F(\mathbf{x}_k, \mathbf{u}_k))
+\alpha h_e(\mathbf{x}_k,\mathbf{e}_k).
\]
Since $\mathbf{u}_k$ satisfies \eqref{eq:u_policy},
\[
\widetilde{\mathcal{C}}\!\left(
-\Delta h_e^{(1:N)}
\right)
\leq0.
\]
Applying \eqref{eq:CPP_ineq} in
Proposition~\ref{p:CPP_Guarantee} with
$f=-\Delta h_e$ gives
\[
\begin{aligned}
\mathbb{P}_{\mathbf{d}^{(1:N)}}\!\Bigg(
\mathbb{P}_{\mathbf{d}_k}\!\Big(
\Delta h_e(\mathbf{x}_k,\mathbf{e}_k,\mathbf{u}_k,\mathbf{d}_k)
&\geq0
\,\Big|\,\mathbf{d}^{(1:N)}
\Big) \\
&\geq1-\delta
\Bigg)
\geq1-\beta,
\end{aligned}
\]
which is exactly \eqref{eq:PECBF_uineq}.
Proposition~\ref{p:PECBF} then gives the stated finite-horizon safety
guarantee.
\hfill$\square$

\begin{remark}
The calibration sample size $N$ controls the magnitude of the
Hoeffding correction $\sqrt{\ln(1/\beta)/(2N)}$, which decreases
as $N^{-1/2}$ for fixed $\beta$.
For $\delta$ and $\beta$ chosen according to
\eqref{eq:PECBF_ineq}, a sufficiently large calibration dataset
is therefore required to satisfy
$\sqrt{\ln(1/\beta)/(2N)}<\delta$.
Increasing $N$ also brings the corrected quantile level closer
to $1-\delta$, potentially reducing the additional conservatism
introduced by the confidence correction.
\end{remark}

Under the conditions of Proposition~\ref{p:last}, enforcing
\eqref{eq:u_policy} at each time step guarantees
\eqref{eq:finite_horizon_safety_goal} with confidence
$1-\widetilde{\beta}$.
A safety filter is employed to minimize the squared Euclidean
deviation
$\|\mathbf u_k-\mathbf k(\mathbf x_k)\|_2^2$
from the nominal input, subject to
$\mathbf u_k\in\mathcal U$ and \eqref{eq:u_policy},
where $\mathbf k:\mathcal X\to\mathcal U$ denotes the nominal
control policy.
The resulting problem is a QP when \eqref{eq:u_policy}
is affine in $\mathbf u_k$ and $\mathcal U$ is defined
by linear constraints.
Algorithm~\ref{alg:CPP_ECBF} summarizes the procedure.

\begin{algorithm}[t]
\caption{CPP-ECBF Safety Filter}
\label{alg:CPP_ECBF}
\begin{algorithmic}[1]

\Statex \textbf{Offline calibration}
\State Obtain calibration samples
$\mathbf{d}^{(1:N)}
\overset{\mathrm{i.i.d.}}{\sim}P_{\mathbf d}$

\Statex
\Statex \textbf{Online control at each $k=0,1,\ldots$}

\State Observe $\mathbf{x}_k$ and $\mathbf{e}_k$

\State Evaluate
$\psi_k^{(i)}
\gets
\psi\!\left(F_e(\mathbf{e}_k,\mathbf{d}^{(i)})\right)$,
$i=1,\ldots,N$

\State Compute
$\widetilde{\mathcal C}
\!\left(\psi_k^{(1:N)}\right)$

\State Compute $\mathbf{u}_k$ as
\Statex \hspace{\algorithmicindent}
$\displaystyle
\mathbf{u}_k^*
\gets
\arg\min_{\mathbf{u_k}\in\mathcal{U}}
\left\|
\mathbf{u_k}-\mathbf{k}(\mathbf{x}_k)
\right\|_2^2$
\Statex \hspace{\algorithmicindent}
$\displaystyle
\quad \text{s.t.}\quad
\phi\!\left(F(\mathbf{x}_k,\mathbf{u_k})\right)
\geq
\widetilde{\mathcal C}
\!\left(\psi_k^{(1:N)}\right)
+
\alpha h_e(\mathbf{x}_k,\mathbf{e}_k)$

\State Apply $\mathbf{u}_k^*$

\end{algorithmic}
\end{algorithm}
\section{Case Study: Adaptive Cruise Control}

This section applies CPP-ECBF to an ACC case study.
The lead-vehicle acceleration is modeled as i.i.d.\ environmental
uncertainty, consistent with the assumption in Section~III.
Monte Carlo simulations are used to empirically assess the
finite-horizon safety guarantee in Proposition~\ref{p:last}.

\subsection{Evaluation Metric}

To quantify the calibration-conditional finite-horizon safety
performance, $I$ independent Monte Carlo trials were conducted,
each with $J$ time steps.
For each trial, a new calibration dataset $d^{(1:N)}$
was sampled and held fixed throughout the trial.
For the $n$th trial, $n=1,\ldots,I$, let
$\mathbf{x}_k^{(n)}$ and $\mathbf{e}_k^{(n)}$
denote the ego-vehicle and environmental states at time $k$,
respectively.
The Conditional Empirical Coverage (CEC) for the $n$th trial
is defined as
\begin{equation}
\mathrm{CEC}_n :=
\frac{
\displaystyle
\sum_{k=1}^{J-H}
\mathbf{1}\!\left(
\bigcap_{i=0}^{H}
\left\{
(\mathbf{x}_{k+i}^{(n)},\mathbf{e}_{k+i}^{(n)})
\in\mathcal{S}_e
\right\}
\right)
}{
\displaystyle
\sum_{k=1}^{J-H}
\mathbf{1}\!\left(
(\mathbf{x}_k^{(n)},\mathbf{e}_k^{(n)})
\in\mathcal{S}_e
\right)
},
\label{eq:CEC}
\end{equation}
where $\mathbf{1}(\cdot)$ denotes the indicator function.
The denominator counts windows starting from a safe state,
whereas the numerator counts those that remain safe for the
following $H$ steps, thereby measuring the observed
finite-horizon safety performance for the fixed calibration
dataset of trial $n$.
The calibration-conditional guarantee was assessed empirically
by checking whether the fraction of trials satisfying
$\mathrm{CEC}_n\geq1-\epsilon$ was at least the prescribed
confidence $1-\widetilde{\beta}$:
\begin{equation}
\frac{1}{I}
\sum_{n=1}^{I}
\mathbf{1}\!\left(
\mathrm{CEC}_n\geq1-\epsilon
\right)
\geq
1-\widetilde{\beta}.
\label{eq:empirical_confidence}
\end{equation}
\subsection{System and Safety Function}
\begin{figure}[t]
  \centering
  \includegraphics[width=\linewidth]{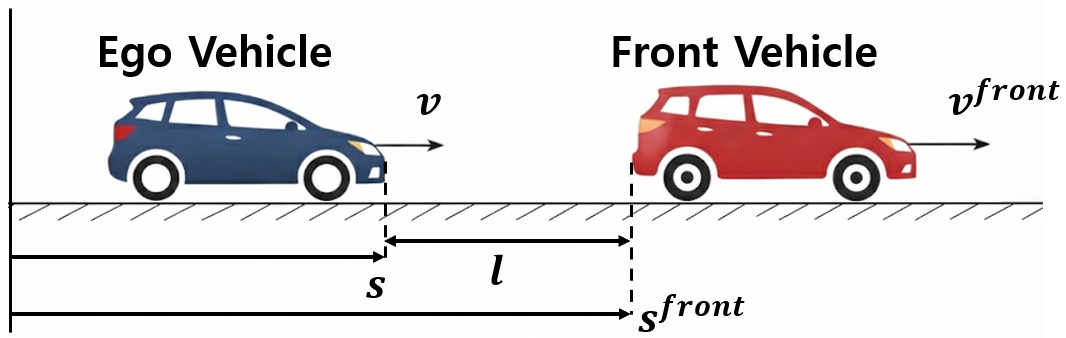}
  \caption{Adaptive cruise control setup and notation.}
  \label{fig:ACC}
\end{figure}
The longitudinal ACC setup is illustrated in
Fig.~\ref{fig:ACC}. Here, $s_k$ and $v_k$ denote the position of
the front bumper and the velocity of the ego vehicle, respectively,
and $s_k^{\mathrm{front}}$ and $v_k^{\mathrm{front}}$ denote
the position of the rear bumper and the velocity of the lead
vehicle, respectively.
Assuming constant acceleration over each sampling interval
$[k\Delta t,(k+1)\Delta t]$, the ego-vehicle and lead-vehicle dynamics
are given by
\begin{equation}
\mathbf{x}_{k+1}
=
\mathbf{x}_k+
\begin{bmatrix}
v_k\Delta t+\frac{1}{2}u_k\Delta t^2\\
u_k\Delta t
\end{bmatrix},
\label{eq:ACC_x}
\end{equation}
\begin{equation}
\mathbf{e}_{k+1}
=
\mathbf{e}_k+
\begin{bmatrix}
v_k^{\mathrm{front}}\Delta t+\frac{1}{2}d_k\Delta t^2\\
d_k\Delta t
\end{bmatrix},
\label{eq:ACC_e}
\end{equation}
where
$\mathbf{x}_k=[s_k,v_k]^T$,
$\mathbf{e}_k=[s_k^{\mathrm{front}},v_k^{\mathrm{front}}]^T$,
$u_k$ is the ego-vehicle acceleration, and $d_k$ is the
lead-vehicle acceleration. 
The lead-vehicle motion can be treated as an exogenous
environmental process not directly affected by the
ego-vehicle acceleration input.
The calibration samples $d^{(1:N)}$ 
and the lead-vehicle accelerations $d_k$ are assumed to be i.i.d.\
draws from a distribution $P_d$ .
Let us consider the environmental safety function
\begin{equation}
h_e(\mathbf{x}_k,\mathbf{e}_k)
:=
\kappa(s_k^{\mathrm{front}}-s_k-l_{\mathrm{desired}})
+
v_k^{\mathrm{front}}-v_k,
\label{eq:ACC_safetyfunction}
\end{equation}
where $\kappa>0$ is a design parameter and
$l_{\mathrm{desired}}>0$ is the desired inter-vehicle distance
parameter. The separable structure of \eqref{eq:ACC_safetyfunction},
together with the i.i.d.\ sampling assumption, allows us
to apply the CPP-ECBF construction.

\subsection{CPP-ECBF Safety Filter}

The safety function in \eqref{eq:ACC_safetyfunction} can be rewritten as
\[
h_e(\mathbf{x}_k,\mathbf{e}_k)
=
\phi(\mathbf{x}_k)-\psi(\mathbf{e}_k),
\]
where
\[
\phi(\mathbf{x}_k)
=-\kappa s_k-v_k-\kappa l_{\mathrm{desired}}, \ 
\psi(\mathbf{e}_k)
=-\kappa s_k^{\mathrm{front}}-v_k^{\mathrm{front}}. 
\]
%
%
Using \eqref{eq:ACC_x} and \eqref{eq:ACC_e} yields
\[
\phi(\mathbf{x}_{k+1})
=
\phi(\mathbf{x}_k)
-\kappa v_k\Delta t
-\left(1+\frac{\kappa\Delta t}{2}\right)u_k\Delta t,
\]
while, for each calibration sample $d^{(i)}$,
\[
\psi_k^{(i)}
=
\psi(\mathbf{e}_k)
-\kappa v_k^{\mathrm{front}}\Delta t
-\left(1+\frac{\kappa\Delta t}{2}\right)d^{(i)}\Delta t.
\]
Therefore,
\[
\begin{aligned}
\widetilde{\mathcal C}\!\left(\psi_k^{(1:N)}\right)
={}&
\psi(\mathbf{e}_k)
-\kappa v_k^{\mathrm{front}}\Delta t\\
&+
\left(1+\frac{\kappa\Delta t}{2}\right)
\widetilde{\mathcal C}\!\left(-d^{(1:N)}\right)\Delta t.
\end{aligned}
\]
The uncertainty term in $\psi_k^{(i)}$ is a positive
multiple of $-d^{(i)}$. Hence, the ordering is preserved,
and the upper conformal quantile of $\psi_k^{(1:N)}$
is obtained from that of $-d^{(1:N)}$ as shown above.
The quantity $\widetilde{\mathcal C}(-d^{(1:N)})$ is referred to as the \emph{conformal uncertainty bound}.
This bound is independent of the vehicle states and
control input and is computed offline for a fixed
calibration dataset.
Substituting this bound into \eqref{eq:u_policy} gives the
deterministic linear constraint on the ego-vehicle
acceleration
\begin{equation}
u_k
\leq
u_{\max,k}^{\mathrm{CPP\text{-}ECBF}},
\label{eq:ACC_u_CBF}
\end{equation}
where
\begin{equation}
\begin{aligned}
u_{\max,k}^{\mathrm{CPP\text{-}ECBF}}
:={}&
-\widetilde{\mathcal C}\!\left(-d^{(1:N)}\right)
-\frac{\kappa}
{1+\frac{\kappa\Delta t}{2}}
\left(v_k-v_k^{\mathrm{front}}\right)\\
&+
\frac{1-\alpha}
{\Delta t\left(1+\frac{\kappa\Delta t}{2}\right)}
h_e(\mathbf{x}_k,\mathbf{e}_k).
\end{aligned}
\label{eq:ACC_umax}
\end{equation}
\begin{remark}
The conformal uncertainty bound
$\widetilde{\mathcal C}(-d^{(1:N)})$ is the only term in
\eqref{eq:ACC_umax} that explicitly depends on the prescribed
safety probability $1-\epsilon$ and confidence
$1-\widetilde{\beta}$.
Since this bound enters \eqref{eq:ACC_umax} with a negative
sign, a larger uncertainty bound yields a more restrictive
acceleration upper limit.
\end{remark}

For the considered ACC problem, the feasible control-input set is
defined as
$\mathcal{U}
:=
[u_{\min},u_{\max}],$
where $u_{\min}$ and $u_{\max}$ denote the lower and upper
longitudinal acceleration limits, respectively.
Following the safety-filter formulation in Section~III, the ACC
controller is given by
\begin{equation}
\begin{aligned}
u_k^\ast(\mathbf{x}_k,\mathbf{e}_k)
:=
&\arg\min_{u_k\in\mathbb{R}}
\frac{1}{2}|u_k-k(\mathbf{x}_k)|^2\\
\mathrm{s.t.}\quad
&
u_k
\leq
u_{\max,k}^{\mathrm{CPP\text{-}ECBF}},\\
&
u_{\min}\leq u_k\leq u_{\max},
\end{aligned}
\label{eq:ACC_QP}
\end{equation}
where $k(\mathbf{x}_k)$ denotes the nominal controller.
Since the objective is quadratic and all constraints are linear
in $u_k$, \eqref{eq:ACC_QP} is a QP.

\subsection{Results and Discussion}

For the Monte Carlo simulations, the lead-vehicle
acceleration $d_k$ and calibration samples $d^{(1:N)}$
were independently drawn from a Laplace distribution
$\mathcal{L}(0,0.5)$, truncated below
at $d_{\min}=-8\,\mathrm{m/s^2}$.
Neither the distribution nor $d_{\min}$ was provided
to the proposed controller, which used only the
calibration samples.
The acceleration limits were set to
$u_{\min}=-0.3g$ and $u_{\max}=0.3g$
($g=9.81\,\mathrm{m/s^2}$).
The parameters were set to $H=10$, $\epsilon=0.2$, and
$\widetilde{\beta}=0.15$, with $\delta=1-(1-\epsilon)^{1/H}$ and
$\beta=\widetilde{\beta}/H$ chosen according to
\eqref{eq:PECBF_ineq}.
The nominal controller $k(\mathbf{x})=\lambda(v_{\mathrm{desired}}-v)$ was used. The remaining parameters were
$\alpha=0.0001$, $\Delta t=0.01\,\mathrm{s}$,
$v_{\mathrm{desired}}=30\,\mathrm{m/s}$,
$\kappa=2\,\mathrm{s}^{-1}$,
$\lambda=0.5\,\mathrm{s}^{-1}$, and
$l_{\mathrm{desired}}=20\,\mathrm{m}$.
The initial states were
$s_1=0\,\mathrm{m}$, $v_1=20\,\mathrm{m/s}$,
$s_1^{\mathrm{front}}=20\,\mathrm{m}$, and
$v_1^{\mathrm{front}}=20\,\mathrm{m/s}$.
$I=10{,}000$ independent Monte Carlo
trials were performed, each with $J=1000$ time steps and
$N=10{,}000$ calibration samples.

\begin{remark}
If the nominal input already satisfies the CPP-ECBF
constraint, the effect of the safety filter may be
difficult to observe.
Therefore, challenging initial conditions
and a desired velocity were selected. These conditions made the constraint
sufficiently active to evaluate how the safety filter
modified the nominal input as shown in \eqref{eq:ACC_umax}.
\end{remark}

The physical acceleration limits remained inactive throughout
all simulated Monte Carlo trials.
Therefore, the CPP-ECBF upper bound was the only active
constraint whenever the safety filter modified the nominal
input.
Accordingly, throughout the simulations, the solution of \eqref{eq:ACC_QP} satisfied
\[
u_k^\ast
=
\min\left\{
k(\mathbf{x}_k),
u_{\max,k}^{\mathrm{CPP\text{-}ECBF}}
\right\}.
\]

\begin{figure}[t]
    \centering
    \includegraphics[width=\linewidth]
    {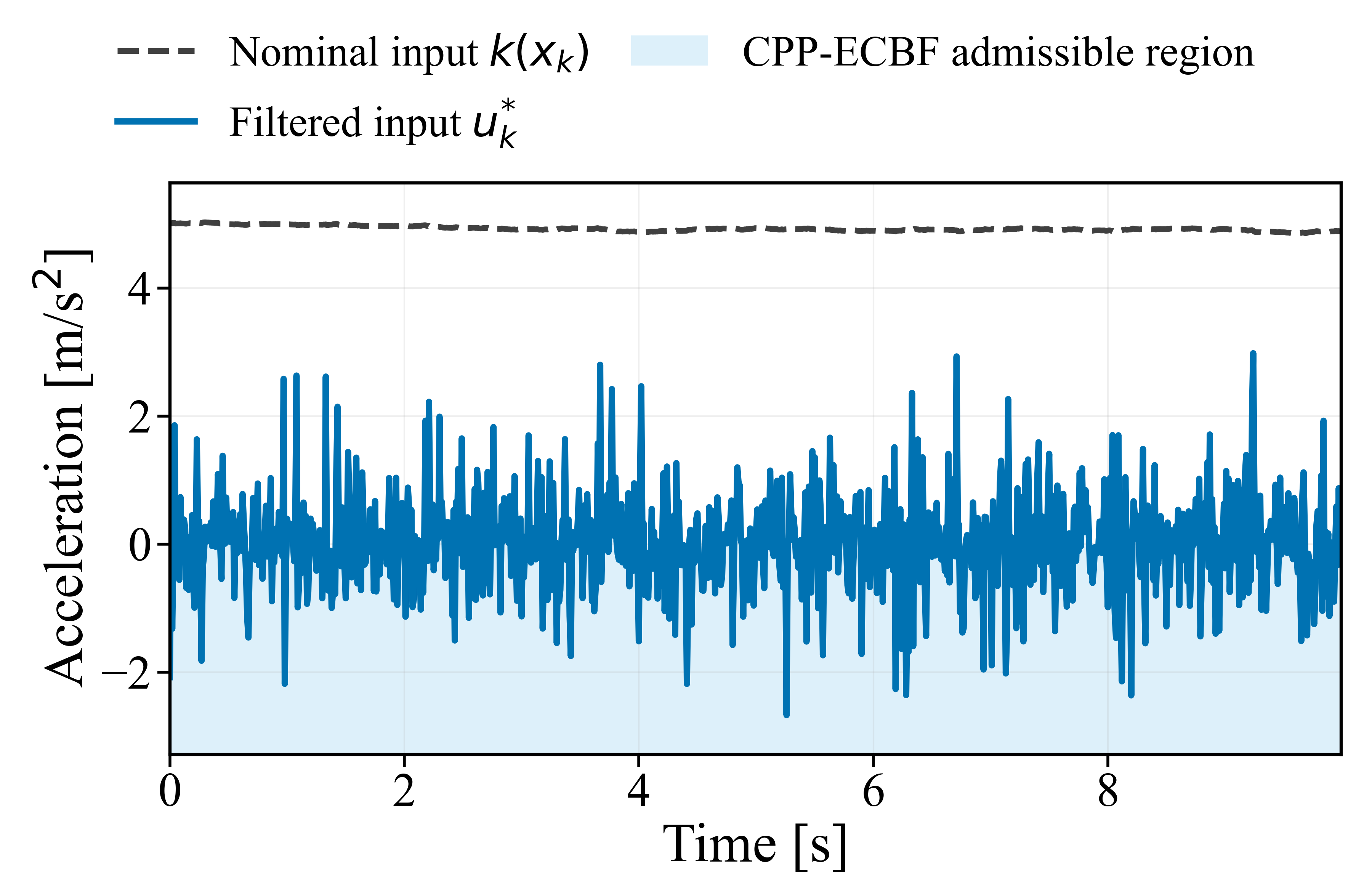}
    \caption{
    Nominal and filtered inputs for the first Monte Carlo
    trial.
    The dark gray dashed line denotes the nominal input
    $k(\mathbf{x}_k)$, the blue solid line denotes the
    filtered input $u_k^\ast$, and the light blue shaded
    region represents the CPP-ECBF admissible region
    $u_k\leq u_{\max,k}^{\mathrm{CPP\text{-}ECBF}}$.
    }
    \label{fig:cpp_ecbf_control}
\end{figure}

Fig.~\ref{fig:cpp_ecbf_control} shows the nominal and filtered
inputs for the first trial of the $I=10{,}000$ Monte Carlo trials.
The nominal input always exceeded the CPP-ECBF acceleration upper
bound by approximately $5\,\mathrm{m/s^2}$ on average because
the desired ego-vehicle speed was substantially higher than
the lead-vehicle speed.
The proposed QP safety filter reduced the nominal input to this bound,
placing the filtered input on the boundary of the shaded
CPP-ECBF admissible input region with minimum deviation
from the nominal input.

Finite-horizon safety was assessed using the CEC metric
in \eqref{eq:CEC} and the empirical confidence criterion
in \eqref{eq:empirical_confidence}.
For CPP-ECBF, $9{,}997$ of $10{,}000$ trials satisfied
$\mathrm{CEC}_n\geq1-\epsilon=0.8$.
The resulting fraction, $0.9997$, exceeded the prescribed
confidence $1-\widetilde{\beta}=0.85$ by $0.1497$.
To examine how much the Hoeffding-based quantile correction
contributed to this difference, CPP-ECBF was compared with
a marginal variant using the standard conformal quantile
$\mathcal C_{1-\delta}(-d^{(1:N)})$.
This marginal quantile omits the Hoeffding-based increase
$\sqrt{\ln(1/\beta)/(2N)}$ and provides coverage at level
$1-\delta$ only in the marginal sense.
It therefore does not provide the calibration-conditional
coverage guarantee required when a single calibration dataset
is held fixed during closed-loop operation.
Hereafter, when distinguishing the two calibration choices
in the following comparison, the CPP-ECBF developed above is
referred to as \emph{conditional CPP-ECBF}, whereas the variant
using the marginal conformal bound is referred to as
\emph{marginal CPP-ECBF}.
For notational convenience, 
let
\[
q_{\mathrm{mar}}
:=
\mathcal C_{1-\delta}\!\left(-d^{(1:N)}\right),
\qquad
q_{\mathrm{cond}}
:=
\widetilde{\mathcal C}\!\left(-d^{(1:N)}\right)
\]
denote the marginal and conditional conformal uncertainty
bounds, respectively.
Each QP safety filter used its corresponding uncertainty bound
in \eqref{eq:ACC_umax} to enforce the acceleration constraint
in \eqref{eq:ACC_u_CBF}.
Within each trial, both methods used the same calibration
dataset and lead-vehicle acceleration sequence, while separate
ego-vehicle trajectories were simulated for the two methods.
\begin{figure}[t]
    \centering

    \begin{subfigure}[t]{\linewidth}
        \centering
        \includegraphics[width=\linewidth]
        {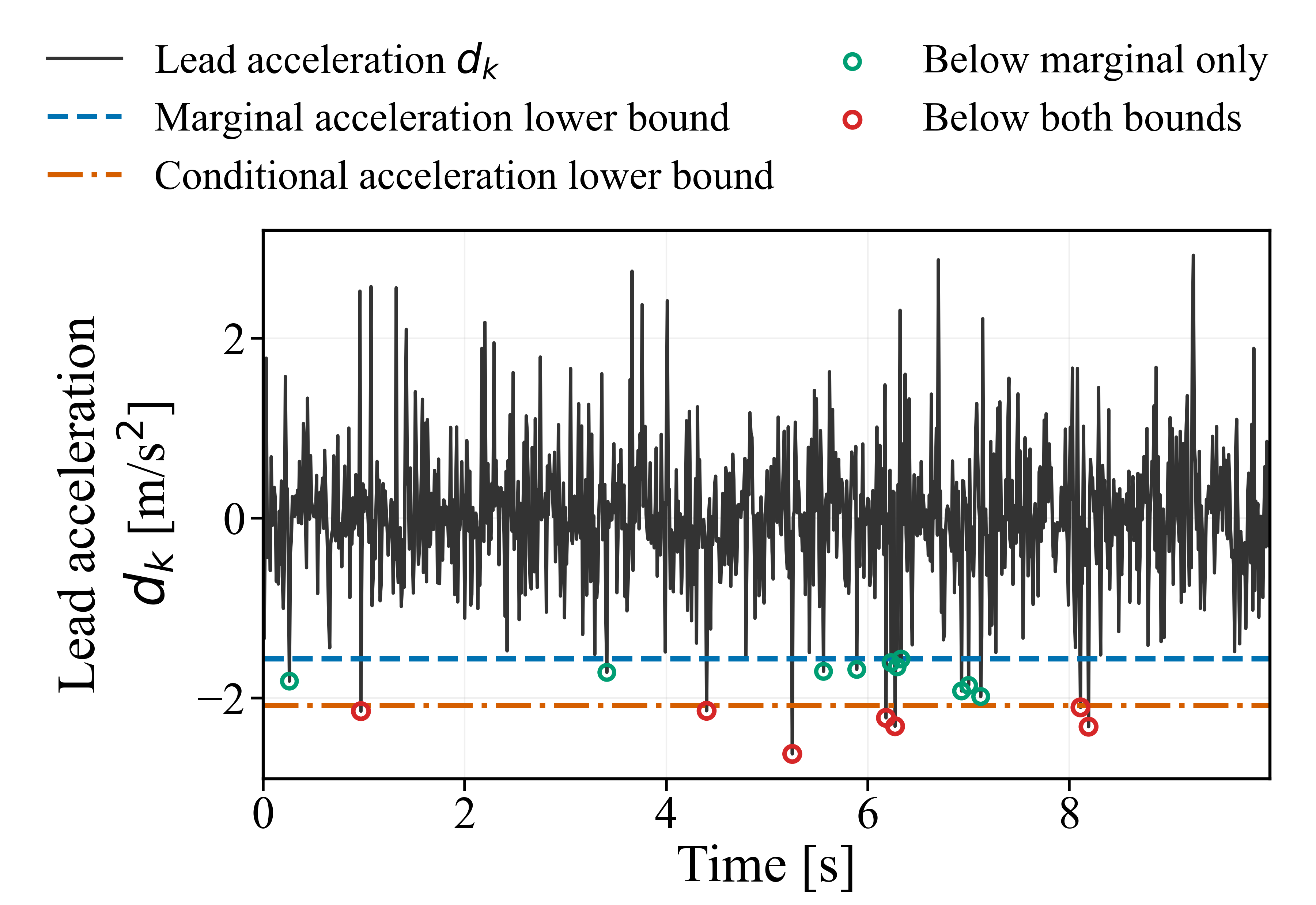}
        \caption{Lead-vehicle acceleration and acceleration lower bounds.}
        \label{fig:lead_acc_bounds}
    \end{subfigure}

    \vspace{1mm}

    \begin{subfigure}[t]{\linewidth}
        \centering
        \includegraphics[width=0.95\linewidth]
        {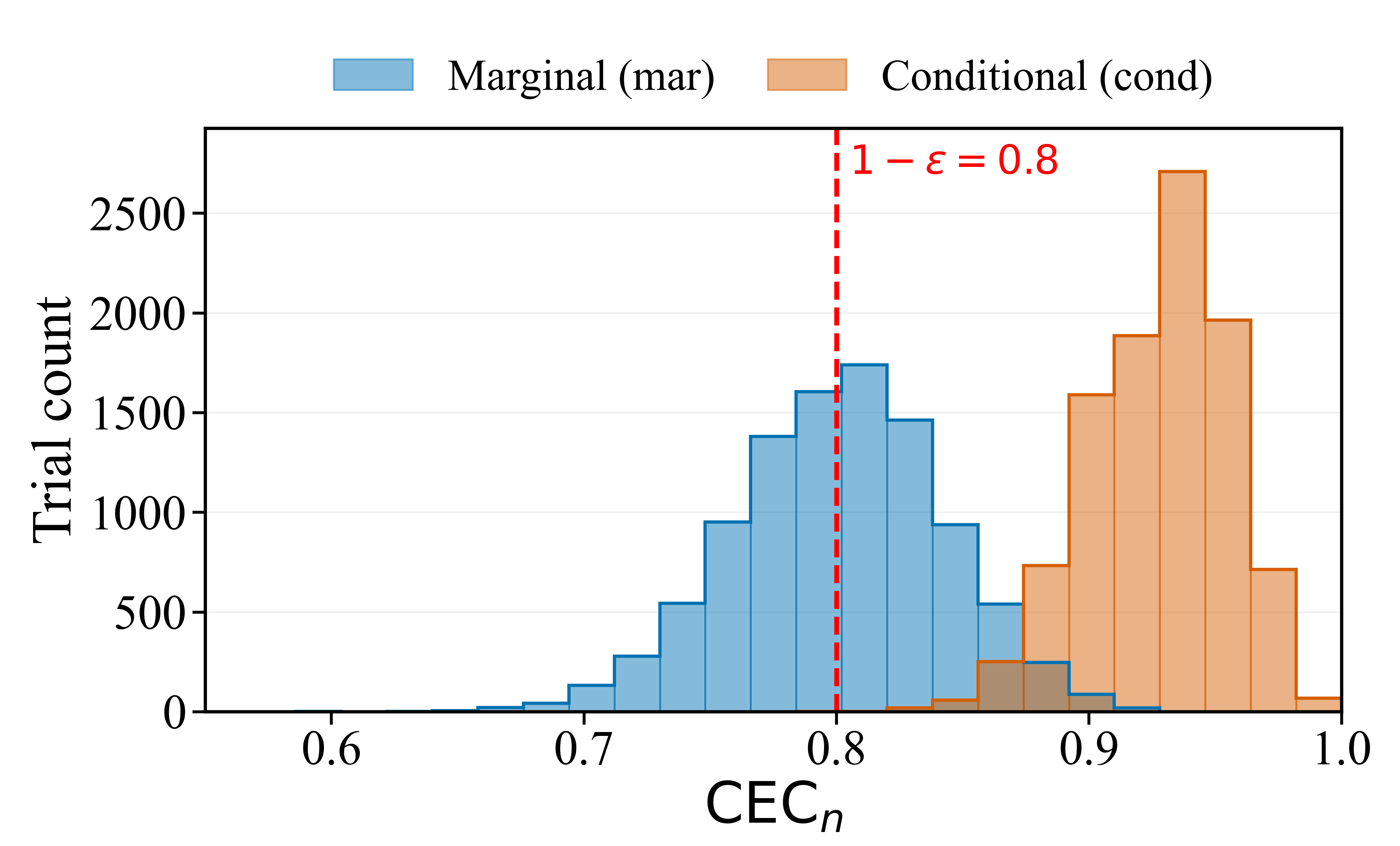}
        \caption{CEC distributions for marginal and conditional CPP-ECBF.}
        \label{fig:cec_comparison}
    \end{subfigure}

    \caption{
    Finite-horizon safety evaluation.
    (a) Lead-vehicle acceleration in the first trial
    (dark gray solid line), together with the marginal lower bound
    $-q_{\mathrm{mar}}$ (blue dashed line) and the conditional lower bound
    $-q_{\mathrm{cond}}$ (orange dash-dotted line).
    Green circles indicate samples below only the marginal lower
    bound, whereas red circles indicate samples below both bounds.
    (b) Overlaid CEC distributions for marginal and conditional
    CPP-ECBF over the Monte Carlo trials.
    The vertical red dashed line indicates $1-\epsilon=0.8$.
    }
    \label{fig:conditional_effect}
\end{figure}

The conformal uncertainty bound $q$ can be interpreted as an
upper prediction bound on $-d_k$ at the corresponding coverage
level. Equivalently,
\[
-d_k\leq q
\quad\Longleftrightarrow\quad
d_k\geq-q,
\]
so $-q$ represents a lower prediction bound on the
lead-vehicle acceleration. When $d_k\geq-q$, the CPP-ECBF acceleration constraint
ensures $\Delta h_e\geq0$.
When $d_k<-q$, the lead vehicle brakes more strongly than
accounted for by the bound, and this inequality is no longer
guaranteed.
When $d_k<-q$, the lead vehicle brakes more strongly than
accounted for by the bound, and this condition is no longer
guaranteed.
Figure~\ref{fig:lead_acc_bounds} illustrates these cases
for the marginal and conditional lower bounds,
$-q_{\mathrm{mar}}$ and $-q_{\mathrm{cond}}$, respectively.
Because the Hoeffding-based correction increases the conformal
quantile level, $q_{\mathrm{cond}}> q_{\mathrm{mar}}$ for
the same calibration dataset, and hence
$-q_{\mathrm{cond}} < -q_{\mathrm{mar}}$.
The green-circled samples satisfy
$-q_{\mathrm{cond}}\leq d_k<-q_{\mathrm{mar}}$.
For these realizations, the conditional constraint ensures
$\Delta h_e\geq0$, whereas the marginal constraint does not
provide this guarantee. The red-circled samples fall below both lower
bounds and are therefore outside the range covered by either
bound. Fig.~\ref{fig:cec_comparison} shows how this difference
is reflected in finite-horizon safety performance over the
Monte Carlo trials. For marginal CPP-ECBF, only $0.5223$ of
the trials satisfied $\mathrm{CEC}_n\geq0.8$, which is below
the prescribed confidence $1-\widetilde{\beta}=0.85$, whereas
conditional CPP-ECBF achieved a trial fraction of $0.9997$.
Accordingly, the empirical CEC distribution is shifted toward
higher values under conditional calibration, consistent with
the additional lead-vehicle braking realizations covered by the
conditional bound in Fig.~\ref{fig:lead_acc_bounds}.

To quantify the corresponding control conservatism, the conformal uncertainty bounds were averaged over the $I=10{,}000$
calibration datasets.
For comparison, the worst-case robust design uses the true
lead-vehicle acceleration lower bound
$d_{\min}=-8\,\mathrm{m/s^2}$, corresponding to
$q_{\mathrm{rob}}=-d_{\min}=8\,\mathrm{m/s^2}$.
This worst-case bound was used only for comparison and was not
provided to the conditional CPP-ECBF controller.
Since the uncertainty bound enters \eqref{eq:ACC_umax} with
coefficient $-1$, differences in the uncertainty bound directly correspond to equal-magnitude differences in the admissible ego-vehicle acceleration upper limit at the same vehicle state.
\begin{table}[t]
    \centering
    \caption{
    Conformal uncertainty bound comparison.
    Marginal and conditional bounds are averaged over
    $10{,}000$ calibration datasets, whereas the
    worst-case robust bound is fixed.
    }
    \label{tab:uncertainty_bounds}
    \small
    \renewcommand{\arraystretch}{1.25}
    \setlength{\tabcolsep}{7pt}
    \begin{tabular*}{\columnwidth}
    {@{\extracolsep{\fill}}lc@{}}
        \hline
        Method
        &
        Conformal uncertainty bound $q$ 
        \\ & on average 
        [$\mathrm{m/s^2}$]
        \\
        \hline
        Marginal CPP-ECBF
        & $1.56$
        \\
        Conditional CPP-ECBF
        & $2.10$
        \\
        Worst-case robust
        & $8$
        \\
        \hline
    \end{tabular*}
\end{table}
As shown in Table~\ref{tab:uncertainty_bounds}, the average
uncertainty bounds for marginal and conditional CPP-ECBF were
$1.56\,\mathrm{m/s^2}$ and $2.10\,\mathrm{m/s^2}$,
respectively.
Thus, conditional calibration reduced the admissible
acceleration upper limit by $0.54\,\mathrm{m/s^2}$ on
average relative to marginal calibration at the same vehicle
state.
This additional conservatism was accompanied by satisfaction
of the prescribed empirical confidence criterion, which the
marginal variant did not achieve.
Compared with worst-case robust enforcement, however,
conditional CPP-ECBF still allowed an acceleration upper limit
that was $5.90\,\mathrm{m/s^2}$ larger on average.

CPP-ECBF constructs the safety constraint using only calibration
samples, without requiring knowledge of the underlying
uncertainty distribution or its support bound.
By allowing a prescribed violation probability rather than
enforcing the constraint against all uncertainty realizations,
the resulting safety filter reduces the conservatism associated
with worst-case robust enforcement while retaining the
prescribed finite-horizon probabilistic safety guarantee under
the stated i.i.d.\ sampling assumptions.
The present guarantee does not directly cover temporal dependence
or distribution shift in lead-vehicle acceleration.
Conformal methods for dependent data and adaptive calibration
could help address these respective limitations.
Moreover, tighter finite-sample calibration rules could reduce
the conservatism of the Hoeffding-based quantile correction.

\section{CONCLUSION}

This paper introduced CPP-ECBF, a distribution-free
sample-based safety-filtering method for environmental
uncertainty with an unknown distribution.
For separable environmental safety functions, the CPP-ECBF
constraint was shown to satisfy the probabilistic ECBF condition
and to provide prescribed finite-horizon safety with confidence
over the calibration dataset under the stated i.i.d.\ sampling
assumptions.
For ACC, the conformal uncertainty bound was computed offline
and yielded an explicit linear acceleration constraint enforced
by a QP safety filter.
In $10{,}000$ Monte Carlo trials, conditional CPP-ECBF satisfied
the empirical finite-horizon safety criterion in $99.97\%$ of
the trials, whereas marginal CPP-ECBF achieved $52.23\%$,
below the prescribed confidence of $85\%$.
At the same time, conditional CPP-ECBF allowed an acceleration
upper limit that was $5.90\,\mathrm{m/s^2}$ larger on average
than worst-case robust enforcement.
Future work will investigate tighter calibration-conditional
bounds, extensions to temporally dependent environmental
uncertainty, and the integration of the derived CPP-ECBF
constraint with broader nominal controller, including
model predictive control and reinforcement-learning-based
controllers.






\bibliographystyle{IEEEtran}
\bibliography{refs}

@article{CBF,
  title={Control Barrier Function Based Quadratic Programs for Safety Critical Systems},
  author={Ames, Aaron D. and Xu, Xiangru and Grizzle, Jessy W. and Tabuada, Paulo},
  journal={IEEE Trans. on Automatic Control},
  volume={62},
  number={8},
  pages={3861--3876},
  year={2017},
  doi={10.1109/TAC.2016.2638961}
}

@inproceedings{CBF_ACC,
  title={Control Barrier Function Based Quadratic Programs with Application to Adaptive Cruise Control},
  author={Ames, Aaron D. and Grizzle, Jessy W. and Tabuada, Paulo},
  booktitle={Proc. IEEE Conf. on Decision and Control},
  pages={6271--6278},
  year={2014}
}

@article{ECBF,
  title={Safety-Critical Control With Input Delay in Dynamic Environment},
  author={Molnar, Tamas G. and Kiss, Adam K. and Ames, Aaron D. and Orosz, Gabor},
  journal={IEEE Trans. on Control Systems Technology},
  volume={31},
  number={4},
  pages={1507--1520},
  year={2023},
  doi={10.1109/TCST.2022.3227451}
}

@inproceedings{PCBF,
  title={Probabilistic Control Barrier Functions: Safety in Probability for Discrete-Time Stochastic Systems},
  author={Mestres, Pol and Werner, Blake and Cosner, Ryan K. and Ames, Aaron D.},
  booktitle={Proc. American Control Conf.},
  pages={1942--1949},
  year={2026}
}

@article{CP_tutorial,
  title={Conformal Prediction: A Gentle Introduction},
  author={Angelopoulos, Anastasios N. and Bates, Stephen},
  journal={Foundations and Trends in Machine Learning},
  volume={16},
  number={4},
  pages={494--591},
  year={2023},
  doi={10.1561/2200000101}
}

@inproceedings{CP,
  title={Conditional Validity of Inductive Conformal Predictors},
  author={Vovk, Vladimir},
  booktitle={Proc. Asian Conf. on Machine Learning},
  pages={475--490},
  year={2012}
}

@article{CPP,
  title={Conformal Predictive Programming for Chance Constrained Optimization},
  author={Zhao, Yiqi and Yu, Xinyi and Sesia, Matteo and Deshmukh, Jyotirmoy V. and Lindemann, Lars},
  journal={arXiv:2402.07407},
  year={2024},
  note={Version 3, July 2026}
}

@article{SA,
  title={The Exact Feasibility of Randomized Solutions of Uncertain Convex Programs},
  author={Campi, Marco C. and Garatti, Simone},
  journal={SIAM J. on Optimization},
  volume={19},
  number={3},
  pages={1211--1230},
  year={2008}
}

@article{SAA,
  title={Sample Average Approximation Method for Chance Constrained Programming: Theory and Applications},
  author={Pagnoncelli, Bernardo K. and Ahmed, Shabbir and Shapiro, Alexander},
  journal={J. of Optimization Theory and Applications},
  volume={142},
  number={2},
  pages={399--416},
  year={2009},
  doi={10.1007/s10957-009-9523-6}
}

@article{CP_MPC,
  title={Safe Planning in Dynamic Environments Using Conformal Prediction},
  author={Lindemann, Lars and Cleaveland, Matthew and Shim, Gihyun and Pappas, George J.},
  journal={IEEE Robotics and Automation Letters},
  volume={8},
  number={8},
  pages={5116--5123},
  year={2023},
  doi={10.1109/LRA.2023.3292071}
}

@inproceedings{CP_Perception,
  title={Safe Perception-Based Control Under Stochastic Sensor Uncertainty Using Conformal Prediction},
  author={Yang, Shuo and Pappas, George J. and Mangharam, Rahul and Lindemann, Lars},
  booktitle={Proc. IEEE Conf. on Decision and Control},
  pages={6072--6078},
  year={2023},
  doi={10.1109/CDC49753.2023.10384075}
}

\end{document}